\documentclass[11pt,letterpaper]{article}
\pdfoutput=1

\usepackage{graphicx}
\usepackage{url}
\usepackage{fullpage}
\usepackage{hyperref}
\usepackage{listings}

\lstdefinestyle{codeStyle}
{
	language=Java,
	frame=single,  
	basicstyle=\scriptsize,
	captionpos=b,
	showstringspaces=false,
	showspaces=false,
	extendedchars=true,
	linewidth=1\linewidth,
	breaklines=true,
	float=htb!  
}

\newcommand{\xf}[1]{Figure~\ref{#1}}

\newcommand{\rpc}{{RPC\index{RPC}}}
\newcommand{\rmi}{{RMI\index{RMI}}}

\newcommand{\jini}{{Jini\index{Jini}}}

\newcommand{\gipc}{{GIPC\index{GIPC}}}

\newcommand{\gee}{{GEE\index{GEE}}}
\newcommand{\geer}{{GEER\index{GEER}}}
\newcommand{\gipsy}{{GIPSY\index{GIPSY}}}

\newcommand{\dms}{{DMS\index{DMS}}}
\newcommand{\dmf}{{DMF\index{DMF}}}

\newcommand{\glu}{{GLU\index{GLU}}}

\newcommand{\gipl}{{GIPL\index{GIPL}}}

\newcommand{\lucid}{{Lucid\index{Lucid}}}

\newcommand{\file}[1]{\url{#1}\index{Files!#1}}

\newcommand{\api}[1]{\texttt{#1}\index{API!#1}}

\newcommand{\lucidL}[1]{{$\mathit{Lucid}$}($L$) }

		{}

\def\myvert{\raise 2.27pt \hbox{\vrule depth 0pt height 8pt width 0.2mm}}
\def\myarrow{\hspace*{0.43mm}%
             \raise 2.29pt\hbox{\vrule depth 0pt height 8pt width 0.16mm}%
             \hspace*{-0.32mm}%
             $\longrightarrow$
             \ %
             }

\begin{document}

\title{Advances in the Design and Implementation of a Multi-Tier Architecture in the GIPSY Environment}

\author{
Bin Han and Serguei A. Mokhov and Joey Paquet\\
Department of Computer Science and Software Engineering\\
Concordia University, Montreal, Quebec, Canada\\
Email: \url{{bin_ha,mokhov,paquet}@cse.concordia.ca}}

\date{}

\maketitle

\begin{abstract}
We present advances in the software engineering design
and implementation of the multi-tier run-time system
for the General Intensional Programming System ({\gipsy})
by further unifying the distributed technologies
used to implement the Demand Migration Framework ({\dmf})
in order to streamline distributed execution
of hybrid intensional-imperative programs using Java.\\\\
{\bf Keywords:} intensional programming,
run-time system,
multi-tier architecture,
General Intensional Programming System ({\gipsy}),
General Eduction Engine ({\gee}),
Demand Migration Framework ({\dmf}),
Demand Migration System ({\dms}),
Jini,
JMS,
multi-threading,
RMI,
Abstract Factory pattern,
Factory Method pattern,
Strategy pattern,
Singleton pattern
\end{abstract}

\maketitle

\section{Introduction}

Intensional programming implies a declarative programming
language based on the denotational semantics~\cite{r-for-semantics-82}.
The declarations are evaluated in an inherent multi-dimensional context
space~\cite{tongxinmcthesis08}. The {\gipsy} project~\cite{gipsy2005,bolu04,mokhovmcthesis05}
aims at providing a platform for the investigation on the intensional and hybrid
intensional-imperative programming. The {\gipsy}'s compiler, {\gipc}, is based
on the notion of Generic Intensional Programming Language
({\gipl})~\cite{paquetThesis,gipsy-simple-context-calculus-08,eager-translucid-secasa08,multithreaded-translucid-secasa08},
which is the core run-time language into which all other flavors of the {\lucid}
(a family of intensional programming languages) language can be translated to.
The notion of a generic language also solved the problem of language-independence
of the run-time system by allowing a common representation for all compiled programs,
the Generic Eduction Engine Resources ({\geer}).
A generic distributed run-time system has been proposed in~\cite{gipsy-multi-tier-sac09},
this paper will present the design and implementation progress so far and the immediate
future work.

\subsection{Problem Statement}

Due to {\lucid}'s denotational (prescriptive) semantics, its inventors had mentioned
the inherent parallelism of {\lucid} programs: ``... the whole programs can be understood as producer-consumer networks computing in parallel. Furthermore, this operational interpretation can be used as the basis of a distributed implementation ...''~\cite{r-for-semantics-82}.

Although a multi-threaded and distributed architecture using Java {\rmi}~\cite{java-rmi}
has been initially designed~\cite{bolu03}, it was not fully integrated
and many of the detailed working flow needed to be clarified. Furthermore,
two more separate branches of distributed computation for {\gipsy}
emerged -- the implementations based on {\jini}~\cite{jini} and
JMS~\cite{jms} of the Demand Migration Framework ({\dmf}), which in
themselves are not interoperable and their top interfaces are not
exactly the same complicating the integration and unification effort
thereby delaying the true parallel or distributed execution of {\lucid}
programs in the {\gipsy}'s implementation of the run-time system -- the
General Eduction Engine ({\gee}).

\subsection{Proposed Solution}

Our work follows upon and enhances on {\glu}'s generator-worker
architecture~\cite{dodd96,glu2,eductive-interpreter} extended
to be multi-tier over the course of multiple design iterations with Java.
We apply most of the high-level design work produced by
Paquet~\cite{gipsy-multi-tier-sac09} by constructing
wrapper classes for each tier type introduced in there, specifically
DGT (Demand Generator Tier),
DST (Demand Store Tier),
DWT (Demand Worker Tier), and the
GMT (General Manager Tier).
Every single GIPSY node in the said design, which usually translates to a single
physical computer, that has been registered within the current
GIPSY network of nodes participating in computation, can host
arbitrary number of instances of each tier.
Since four local and distributed computation prototypes
are implemented, which are multi-threaded and
RMI~\cite{bolu03,mokhovmcthesis05},
Jini~\cite{vassev-mscthesis-05}, and JMS~\cite{pourteymourmcthesis08}
we decided to integrate them together by applying the abstract factory,
factory method and strategy design patterns~\cite{soen-design-patterns-1995}
following extreme programming~\cite{xtreme-programming-explained-2005}
and the test-driven development~\cite{junit,mokhovmcthesis05,tongxinmcthesis08}
methodologies aiming at constructing a framework with high extensibility
and maintainability for the further iterations.

\subsection{Related Work}

The work presented in this paper is an evolution of the
original architecture for the run-time system of the {\gipsy},
as hinted in~\cite{paquetThesis}, and briefly presented
in~\cite{gipsy-arch-2000}. The architecture proposed in
these works was itself developed following the
generator-worker architecture adopted successfully by {\glu}~\cite{glu1,glu2}.
Despite {\glu}'s successful implementation, the run-time system of {\glu}
was not as scalable and flexible as the solution that we are designing for and
implementing in this paper.
In addition, the communication procedures of the run-time system
implemented in {\glu} was implemented using {\rpc}.
Our solution proposes a much more flexible approach by integrating
demand migration and storage by using the Demand Migration Framework ({\dmf}),
which can be concretely instantiated using various middleware technologies,
such as {\jini}~\cite{dmf-plc05,dmf-cnsr05} and JMS~\cite{dmf-pdpta07,dms-pdpta08}.

\subsubsection{Node and Tier Properties}

In a {\gipsy} peer-to-peer computing network of nodes and tiers
we aim for the following properties:

\begin{itemize}
\item
Demands are propagated without knowing where they will
be processed or stored.

\item
Any tier or node can fail without the system to be
fatally affected.

\item
Nodes and tiers can seamlessly be added or
removed on the fly as computation is happening.

\item
Nodes and tiers can be affected at run-time
to the execution of any GIPSY program,
i.e. a specific node or tier could be
computing demands for different programs.

\end{itemize}

The conceptual design of a GIPSY node is in \xf{fig:GIPSYnodeDesign}.

\begin{figure}[htb!]
	\begin{centering}
	\includegraphics[width=.47\textwidth]{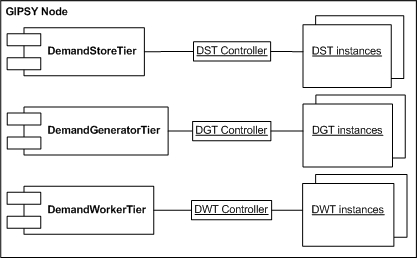}
	\caption{Design of the GIPSY Node}
	\label{fig:GIPSYnodeDesign}
	\end{centering}
\end{figure}

\subsubsection{Demand Driven Computation}

The central concept to this model of execution
is the notion of generation, propagation, and computation
of demands and their resulting values. We have demands
of several types: {\em intensional}, {\em procedural},
{\em resource}, and {\em system}. Intensional demands are of the form:

\begin{center}
\begin{scriptsize}
\noindent{\tt \{GEERid, programId, context\}}
\end{scriptsize}
\end{center}

\noindent
where \api{GEERid} is a unique identifier for the {\geer}
(i.e. the compiled program) that this demand was generated for;
\api{programId} is an identifier declared in this {\geer}
(in this case a Lucid identifier);
and \api{context} is the context of evaluation of this demand. Procedural demands are of the form:

\begin{center}
\begin{scriptsize}
\noindent{\tt\{GEERid, programId, Object params[], context, [code]\}}
\end{scriptsize}
\end{center}

\noindent
where \api{GEERid} is a unique identifier for the {\geer}
that this demand was generated for; \api{programId} is an
identifier declared in this {\geer} (in this case a procedure identifier); 
\api{params[]} is an array of \api{Object}s that this procedure takes as
arguments, \api{context} is the context of evaluation of this demand,
and {\tt [code]} is the (optional) executable code of the procedure,
in cases where we don't want to assume that the worker has the code
to be executed available. Resource demands are of the form:

\begin{center}
\begin{scriptsize}
\noindent{\tt\{resourceTypeId, resourceId\}}
\end{scriptsize}
\end{center}

\noindent
where \api{resourceTypeId} is an identifier for a
resource type, which is an enumerated type now containing
a {\geer} and possibly a procedure class.
This enumerated type is extensible in order to allow
for new resource types to be added later.
The \api{resourceId} is the unique identifier
for the specific resource instance being sought for
by the demander (e.g. demand generator). Any new resource
type created is provided with a unique identifier scheme
to identify each specific resource instance of this type. System demands are of the form:

\begin{center}
\begin{scriptsize}
\noindent{\tt\{destinationTierId, systemDemandTypeId, Object params[]\}}
\end{scriptsize}
\end{center}

\noindent
where \api{destinationTierId} is the tier unique identifier of the tier
to which this demand is addressed, \api{systemDemandTypeId} is an identifier
for a system demand type, which is an enumerated type containing
one element for each kind of system demand, and \api{params[]}
is an array of \api{Object}s that this system demand
takes as arguments, if any.

\subsubsection{Demand Identifiers}

\paragraph*{Universally unique identifier}

The \api{DispatcherEntry} class uses a universally unique identifier
to uniquely identify each demand within a computing GIPSY network.
This identifier is local to the {\dms} and is invisible from the
perspective of GIPSY tiers. Note that this identifier scheme
generates distinct identifiers for absolutely all demands produced
in the system.

\paragraph*{Demand signature identifier}

This is the identifier of a demand that is generated from its
tuple elements. Note that all demands generated with the same
signature (e.g. {\tt\{GEER1,A,[d:2]\}}) will generate the same
demand signature identifier, so that the same demand generated
after having been computed can be queried in the DST and
its result can be extracted without recomputation, following
the principles of dynamic programming.

\subsubsection{Multi-Threaded and RMI Run-Time}

Single-threaded, multi-threaded, and {\rmi}-based
run-time architectures with the NetCDF-supported
storage for {\gipsy} have been explored
first~\cite{bolu03,leitao04,mokhovmcthesis05,netcdf}
and an attempt of unification of them has been made.
The requirement of the new multi-tier architecture
states that they must be a possibility still
when needed in addition to the new technologies
that were further developed that are discussed
in the following section.

\subsubsection{Jini-DMS and JMS-DMS}

To overcome {\glu}'s inflexibility, the {\gipsy} was designed to include a generic and
technology-independent collection of Demand Migration Systems ({\dms}s),
which implement the Demand Migration Framework ({\dmf}) for a particular
technology or technologies to communicate and store information.
Jini-DMS~\cite{vassev-mscthesis-05} incorporates a solution
based on {\jini}~\cite{jini} and JavaSpaces~\cite{javaspaces,jinijavaspace}, 
where {\jini} has been used for the design and implementation of
the Transport Agents (TAs) and JavaSpaces for the design and
implementation of the Demand Store.
JMS-DMS~\cite{pourteymourmcthesis08} applied the {\dmf} framework
based on the Java Messaging Service (JMS) paradigm.
The JBoss Application Server~\cite{jbossguid} has been used
as JMS provider and Hypersonic Database (HSQLDB)~\cite{hsqldb},
which is an embedded solution inside JBoss provides
persistence and caching.

\section{Design and Implementation}

We present the detailed design decisions so far
of our ongoing implementation here.

\subsection{Multi-tier Package}

Classes and interfaces for the implementation under the
\api{gipsy.GEE.multitier} package. The corresponding wrappers
are located in their respective subdirectories (sub packages).
To summarize, we have a root \api{multitier} package:
\begin{verbatim}
gipsy.GEE.multitier
\end{verbatim}
In there there are subpackages for each of the tier types and the
corresponding wrapper classes among other things:
\begin{verbatim}
gipsy.GEE.multitier.DGT.DGTWrapper
gipsy.GEE.multitier.DST.DSTWrapper
gipsy.GEE.multitier.DWT.DWTWrapper
gipsy.GEE.multitier.GMT.GMTWrapper
\end{verbatim}

\subsection{Wrappers, API and Classes}

Regarding the core design of the Multi-Tier Architecture and
the developers' implementation
efficiency, we decided to define the four aforementioned wrapper classes:
\api{DGTWrapper} (Demand Generator Tier Wrapper),
\api{DSTWrapper} (Demand Store Tier Wrapper),
\api{DWTWrapper} (Demand Worker Tier Wrapper), 
and
the \api{GMTWrapper} (General Manager Tier Wrapper), such that
they all inherit from the same abstract class called
\api{GenericTierWrapper} that implements the most common functionality
of the interface \api{IMultiTierWrapper}.

The interface is to be used by any invoking
application, e.g. the main class \api{GEE}
or even some eventual external applications.
We defined the preliminary API of the interface, and
we will provide the code for it and adjust the tier wrapper stubs to
adhere to the interface.

The interface we designed, is placed into the same package \api{gipsy.GEE.multitier}
and is called \api{IMultiTierWrapper}. It is used by the \api{GEE} main
class or the tier controller classes to hold a reference to a given tier type
in a generic manner.
The initial content of the interface is in \xf{list:multi-tier-wrapper-interface-api}
(trimmed) that the above actual wrapper classes implement.
Thus, we define the actual Java
syntax interface and its implementation within the concrete wrapper
classes.

\begin{lstlisting}[
    label={list:multi-tier-wrapper-interface-api},
    caption={Primary API of the \api{IMultiTierWrapper} Interface.},
    style=codeStyle
    ]
import gipsy.Configuration;

public interface IMultiTierWrapper
extends Runnable
{
    ...
    startTier();
    stopTier();
    setConfiguration(Configuration);
    Configuration getConfiguration();
    ...
}
\end{lstlisting}

\subsection{Generic Wrapper}

Two GIPSY objects included as data members in the \api{GenericTierWrapper} adhere to the APIs of
\api{Configuration} and \api{ITransportAgent}, and are described further.
All wrappers are to have a configuration instance and potentially communicate
through a given transport agent (TA) implementation.

\begin{itemize}

\item
\api{Configuration} contains a \api{Serializable} configuration of this {\gipsy} instance
and its components, for static and run-time configuration management.

\item
A TA reference abstracted by the \api{ITransportAgent} unification interface for all transport agents (TAs)
implemented in the extended DMF (Demand Migration Framework) and DMS
(Demand Migration System) and beyond. All TAs must implement this interface. This is a
super-interface
for the use by the engine and the multi-tier architecture.
The original implementations based on {\jini} and JMS did not
have a common super-interface, which we had to define
and provide ourselves during the course of this work.
After defining a family of interfaces, we can encapsulate
each implementation and make them interchangeable. The strategy pattern
lets the implementation technique vary independently from clients that use it.
\end{itemize}

\subsection{DGT and DWT Wrappers}

Both \api{DGTWrapper} and \api{DWTWrapper} classes
have the common \api{oGEERPool} data member, which is
a collection of objects of type \api{GIPSYProgram}, which
act like local caches of the downloaded programs
from the DST for execution.
These are the brief descriptions of the contained
components:

\begin{itemize}

\item
\api{GIPSYProgram}:
A dictionary of identifiers and the abstract syntax tree (AST)
compiled from the program by the compiler, {\gipc},
also mentioned as a {\geer} (General Eduction Engine Resources).

\item
\api{GEERPool}:
Maintains a collection of {\geer}s, serving as a cache in each DWT
and DGT. Whenever a demand is needed pertaining to an identifier embedded in a particular GEER, DWT and DGT will search for this GEER in
the local GEER pool, and if it's not cached, a resource demand is made in order
to get the required GEER from elsewhere.

\end{itemize}

\subsection{DST Wrapper}

The \api{DSTWrapper} is similar to the two earlier described classes,
except when it inherits from the base class \api{GenericTierWrapper},
the \api{DSTWrapper} encapsulates \api{oStorageSubsystem},
which is an object of type \api{IVWInterface}.

\begin{itemize}

\item
\api{IVWInterface} is an integrated Intensional Value Warehouse, that
now we refer to as a demand store (DS), specifically, if
the Demand Migration System used by the DST is implemented by JMS~\cite{jms},
\api{IVWInterface} represents JBoss~\cite{jbossguid};
if implemented by {\jini}, represents JavaSpaces~\cite{javaspaces},
if locally or by {\rmi}, then represents NetCDF~\cite{netcdf}.

\end{itemize}

\subsection{Supporting Classes}

Throughout our design and development effort we further introduce
a data structure, a factory method class for tiers, and the controller
class to satisfy the high-level design presented earlier.

\begin{itemize}

\item \api{EDMFImplemenation} --
\api{EDMFImplemenation} is an enumerated type containing options for multi-threaded, RMI, Jini and JMS-based
communications middleware so far. It acts as an identifier for the techniques
used in the implementation. In the future, others may add other types
representing the new DMSs they might develop.

\item \api{TierFactory} --
\api{TierFactory} is an instance of the abstract factory pattern.
Each tier controller (see below) maintains a reference to the subclass
of \api{TierFactory}, which creates objects of its respective tier type.
Each concrete tier factory, e.g. \api{DGTFactory}, provides its
controller, e.g. \api{DGTController}, an interface to create families
of DGT without specifying their concrete implementation transport agent
strategies, which enhances the flexibility and maintainability of the system.

\item
\api{NodeController} --
\api{NodeController} is an abstract class for each tier controller. 
Its subclasses decide, which tier type to instantiate.
Its existence facilitates further implementation and addition of any new
tier types to the multi-tier architecture as needed.
If necessary, there will be one controller for each tier that implements
\api{GenericTierWrapper} running on every node, which is a physical
computer. It takes responsibility of adding to (through \api{TierFactory}) or
removing any tier instances from the node. The singleton pattern is applied
here to ensure only one controller is created for each tier on each node,
and the factory method pattern lets the subclasses to specify the tier
objects it controls.
\end{itemize}

\subsection{Design Summary}

In \xf{fig:multitier-package} is the UML diagram describing
the relationships between different tier types, the configuration,
TAs, the engine, and other modules we just described. This overall
class diagram was abbreviate to remove the usual routine details
most classes have, focusing only on the core aspects of this work.

\begin{figure*}[htb!]
	\centering
	\includegraphics[width=\textwidth]{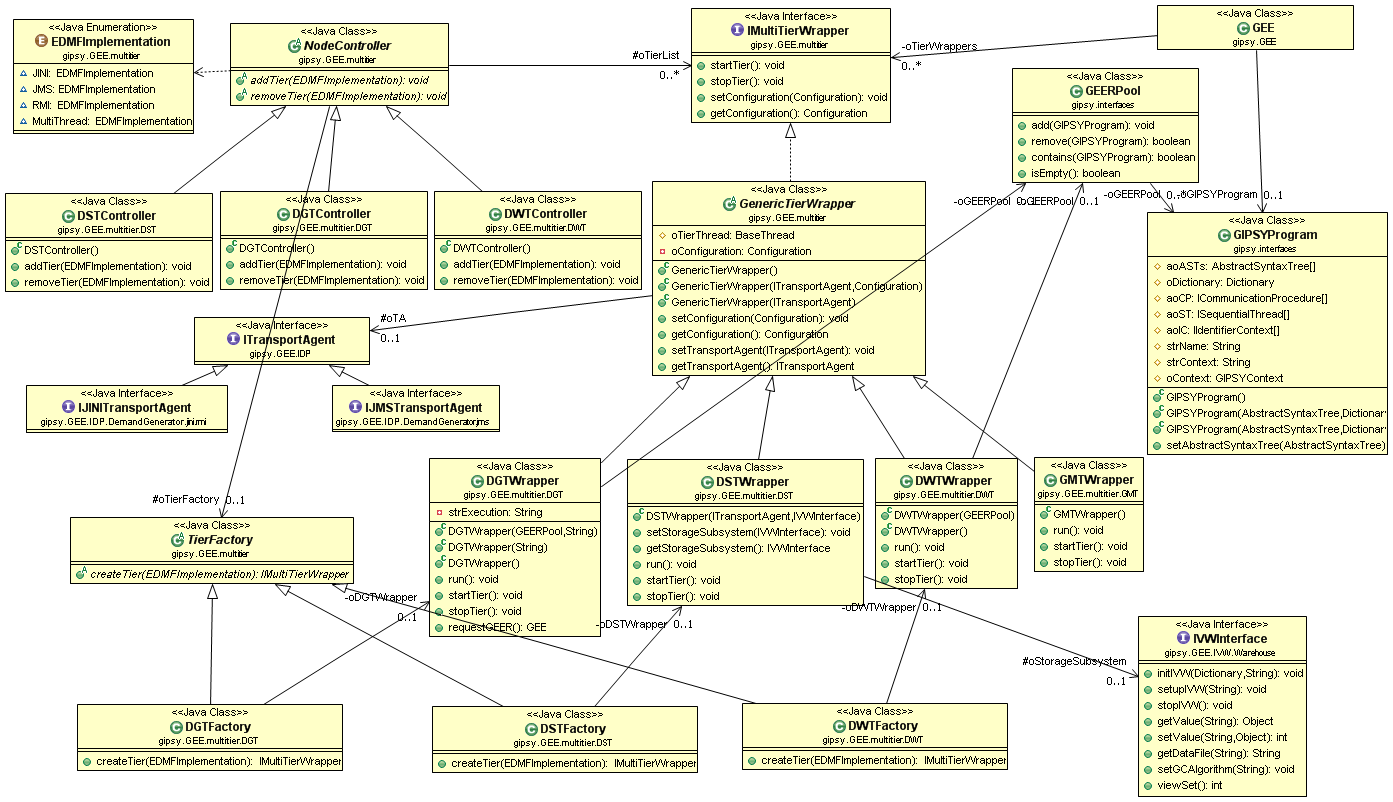}
	\caption{The Initial Multi-tier Architecture Design and Implementation}
	\label{fig:multitier-package}
\end{figure*}

\subsection{Integration with the {\gee}}

The primary integration of the invocation of
the multitier services via the main entry point
of the engine, the main class \api{GEE}, has
to be re-designed to accommodate these
new developments.
The adjustments include option processing,
service start-up and handling the control
over the a particular tier via its controller
or directly for preliminary testing purposes.

For the immediate support of the wrapper invocation
we provide the
means through options of \file{GEE.java} (the main class) to invoke a particular
tier or tiers. We also create the wrapper Linux, MacOS X, and Windows
shell and batch scripts to
start up the tiers,
and the GNU Make~\cite{gmake}'s \file{Makefile}s to build the new multitier code in Linux.

In order to start the {\gee} (that roughly corresponds
to a GIPSY node instance) we do:

\begin{verbatim}
./gee --option
\end{verbatim}

\noindent
e.g. to start up a DGT, it is:

\begin{verbatim}
./gee --dgt[=N]
\end{verbatim}

\noindent
(where N is the optional number of instances) and the corresponding
wrapper scripts, such that one can start any tier, like:

\begin{verbatim}
./dgt N
./dst N
...
\end{verbatim}

{\gee} uses the newly created API in the previous section
of \api{IMultiTierWrapper}. Based on the options, \api{gee}
employs the factory method to instantiate the desired
tier types of arbitrary number of instances, and then
starts them or stops them as directed using the API
of \api{IMultiTierWrapper}. Some of this first build
interaction is under the process of migration to the
\api{NodeController} class described earlier, i.e.
\api{GEE} will delegate the start up activities and
others to the node controller, which will in turn
will use the factory pattern to manage the tiers.
After having started the tiers, \api{GEE} proceeds
to the execution of the program if there is one
to execute, encapsulated into a \api{GIPSYProgram}
instance, the {\geer}. As execution progresses,
the demands are generated and handed off to the
tiers responsible for the delivery and results
computation and return along with the warehouse store caching
principles described earlier.

\section{Conclusion}

We believe that the ongoing design and implementation presented in this work provides
a feasible solution for the eductive evaluation of hybrid
intensional-imperative programs and tier management.
Especially, the multi-tier infrastructure, once fully implemented and tested,
will offer the {\gipsy} run-time system
high scalability and flexibility that was pending integration
effort from various developers for a long time.

We had to add some extra layers of abstraction in terms of
the interfaces and APIs of interfaces \api{IMultiTierWrapper},
\api{ITransportAgent}, and class \api{GEERPool} in order to remain
extensible and flexible to accommodate any future changes
to the design and implementation. We defined a new package
for the multitier implementation, and described the
details and the relationship of the core classes used
in the design.

\section{Future Work}

We have listed the following short-term immediate and longer
term future work items:

\begin{itemize}
\item
Further integration of the multi-tier run-time with the rest of {\gee}.

\item
Extensive unit testing and integration testing.

\item
Performance testing on a hybrid network and cluster environments.

\item
Security layer integration and testing~\cite{marf-gipsy-distributed-ispdc08}.

\end{itemize}

\section{Acknowledgments}

We acknowledge the reviewers of this work
and their constructive feedback.
This work was sponsored in part by NSERC and the
Faculty of Engineering and Computer Science,
Concordia University, Montreal, Canada.

\bibliographystyle{abbrv}
\bibliography{gipsy-multi-tier-impl-arXiv}

\end{document}